\documentclass[12pt]{article}

 \newread\testifexists
 \def\GetIfExists #1 {\immediate\openin\testifexists=#1
     \ifeof\testifexists\immediate\closein\testifexists\else
     \immediate\closein\testifexists\input #1\fi}

 \usepackage{gthstyle}\usepackage{amsfonts}
 \usepackage{amssymb}
 \usepackage{graphicx} \usepackage{epstopdf}
 \immediate\openin\testifexists=e
     \ifeof\testifexists\immediate\closein\testifexists\else
     \immediate\closein\testifexists\input e\fipsf

 \def\Bbb#1{\setbox0=\hbox{$\tt #1$}  \copy0\kern-\wd0\kern .1em\copy0}

 \def\bbf#1{\setbox0=\hbox{$#1$} \kern-.025em\copy0\kern-\wd0
         \kern.05em\copy0\kern-\wd0 \kern-.025em\raise.0433em\box0}

 \immediate\openin\testifexists=a
     \ifeof\testifexists\immediate\closein\testifexists\else
     \immediate\closein\testifexists\input a\fimssym.def  
 
 \newcommand{\tl}[1]{\tilde{#1}}        \renewcommand{\^}[1]{\hat{#1}}      
                     
 \newcommand{\nn}{\nonumber\\[2pt]}             \newcommand{\nm}{\nonumber}
 \newcommand{\be}{\begin{eqnarray}}             \newcommand{\ee}{\end{eqnarray}}
 \newcommand{\bi}[1]{\begin{itemize}\item[#1]}           \newcommand{\ei}{\end{itemize}}
 \newcommand{\eqn}[1]{(\ref{#1})}

 \newcommand{\crlb}[1]{\label{#1}\\[2pt]}
 
 \newcommand{\crld}[1]{\label{#1}}
 \newcommand{\eela}[1]{\quad\hbox{\scriptsize{#1}}\label{#1}\end{eqnarray}}
 \newcommand{\eelb}[1]{\label{#1}\end{eqnarray}}
 
 \newcommand{\newsecb}[2]{\section{#1}\label{#2}\setcounter{equation}{0}}

 \newcommand{\nolabels} {\def\eel{\eelb} \def\crl{\crlb} \def\newsecl{\newsecb}\def\bibiteml{\bibitem}\def\citel{\cite}\def\labell{\crld}}

\newcommand\publishversion{\nolabels\setlength{\textheight}{9in}\setlength{\oddsidemargin}{0in}
    \setlength{\textwidth}{6.3in}\setlength{\topmargin}{-0.1in}}

 \def\a{\alpha}      \def\b{\beta}   \def\g{\gamma}      \def\G{\Gamma}
 \def\d{\delta}         
 \def\k{\kappa}      \def\l{\lambda} \def\L{\Lambda}     \def\m{\mu}
             \def\vv{\varphi}    
 \def\j{\psi}            \def\r{\varrho}     \def\s{\sigma}  
 \def\t{\tau}        \def\th{\theta}  
               
 \def\w{\omega}        

       \def\NN{\mathcal{N}}
 \def\pa{\partial} \def\ra{\rightarrow}
  
 \def\dd{{\rm d}}  \def\bra{\langle}   \def\ket{\rangle}

 \def\iss{\ =\ }

 \def\fract#1#2{{\textstyle{#1\over#2}}}
 \def\ffract#1#2{\raise .2 em\hbox{$\scriptstyle#1$}\kern-.3em/
                 \kern-.2em\lower .15 em \hbox{$\scriptstyle#2$}}
 \def\fractje#1#2{{\scriptstyle{#1\over#2}}}
 \def\half{\fract12}  
 \def\halfje{\fractje12}
 \def\ex#1{e^{\,\textstyle#1}}
  
\def\bmatrix{\begin{matrix}} \def\ematrix{\end{matrix}} \def\bpmatrix{\begin{pmatrix}}\def\epmatrix{\end{pmatrix}}
\def\bcenter{\begin{center}} \def\ecenter{\end{center}}

\def\lowerheightfig#1#2#3{\(\raise-#1\hbox{\includegraphics[height=#2]{#3}}\)}
\def\lowerwidthfig#1#2#3{\(\raise-#1\hbox{\includegraphics[width=#2]{#3}}\)}
	
	\def\inn{{\mathrm{in}}}  \def\out{{\mathrm{out}}}   \def\tot{{\mathrm{tot}}} \def\ds{\displaystyle}

\publishversion
\begin{document} \begin{titlepage}

\title{\normalsize \hfill ITP-UU-15/12  
\vskip 20mm \Large\bf Diagonalizing the\\ Black Hole Information Retrieval Process}

\author{Gerard 't~Hooft}
\date{\normalsize Institute for Theoretical Physics \\
Utrecht University \\ and
\medskip \\ Spinoza Institute \\ Postbox 80.195 \\ 3508 TD Utrecht, the Netherlands \smallskip \\
e-mail: \tt g.thooft@uu.nl \\ internet: \tt
http://www.staff.science.uu.nl/\~{}hooft101/}

\maketitle

\begin{quotation} \noindent {\large\bf Abstract } \medskip \\
{\small The mechanism by which black holes return the absorbed information to the outside world is reconsidered, and described in terms of a set of mutually non-interacting modes. Our mechanism is based on the mostly classical gravitational back-reaction. The diagonalized formalism is particularly useful for further studies of this process. Although no use is made of string theory, our analysis appears to point towards an ensuing string-like interaction. It is shown how black hole entropy can be traced down to classical gravitational back-reaction.}
\end{quotation}

\vfill \flushleft{September 4, 2015}

\end{titlepage}

\eject
\setcounter{page}{2}
\newsecl{Introduction}{intro}
The work described here was inspired by a recent presentation given by S.~Hawking\,\cite{Hawking1} where he presented his modified views on the subject of the black hole information paradox. The approach favoured by the present author was presented earlier in Refs.\,\cite{GtHBH}\,\cite{Polchinski}. We assume the basic mechanism at work here to be the gravitational back reaction of matter entering a black hole onto particles emitted by the Hawking process. As long as momentum exchange in the transverse directions is kept small, the calculation of the back-reaction process is straightforward and unquestionable, even if the outcome forces us to regard the in- and out going stream of particles in an unusual way. The only questionable thing is the transverse momentum cut-off.

We begin with a brief recapitalization of the earlier calculations, noting that the results presented earlier were still not in diagonal form, and this can be further improved. No explicit reference is made to string theory.  For simplicity, we concentrate on a small segment of the black hole horizon, such that it can be represented in terms of Rindler space. Later, one can always substitute that by the full Schwarzschild metric, replacing the transverse Rindler coordinates \(\tl x\) by \((\th,\ \vv)\), and the transverse Rindler momenta \(\tl k\) by the angular momentum quantum numbers \((\ell,\,m)\). Our analysis applies to all non-extreme black hole configurations. The interactions are basically local.

Although it should be clear from the procedure that, written in this form, gravitational back reaction is sufficiently strong to imprint all ingoing information onto the Hawking particles going out, the expressions obtained can be worked out further, to show explicitly how the information flows here. To this end, we divide matter flowing in and out into modes that are mutually independent, acting as if they are bouncing against a brick wall. The difference between our present brick wall and an idea proposed long ago\,\cite{GtHBH0}, is that the present wall was derived from first principles, and it depends on longitudinal and transverse wave numbers. Unfortunately, we were not yet able to reproduce Hawking's entropy expression, since the micro-states still diverge, but it is suspected that this shortcoming can be repaired by inspecting more closely the relationship between our methods and string theories.

\newsecl{The back reaction}{back}	

The gravitational effect of a fast moving particle on the metric is given by the Aichelburg-Sexl solution\,\cite{AichelSexl}. This solution shows a weak singularity in a flat, transverse plane through the moving particle. Write the transverse coordinates as \(\tl x=\big( {x\atop y}\big)\), and the longitudinal ones as \(z^\pm=\fract 1{\sqrt 2}(z\pm t)\). A flat metric is given by \(\dd s^2=\dd\tl x^2+2\dd z^+\dd z^-\). The particle is assumed to be massless.

Before a transverse plane arrives, the Aichelburg Sexl metric is flat in the coordinates \(u^\m_<=(\tl x,\, z^\pm_<)\); after the particle passed, the metric is flat in the coordinates \(u^\m_>=(\tl x,\, z^\pm_>)\), and on the transverse plane these flat coordinates are joined together as
\be u^\m_>=u^\m_<-4Gp^\m\log(|\tl x|/C)\ ,\eel{shift}
where \(p^\m\) is the 4-momentum of the massless particle, \(\tl x\) is the transverse distance from the particle,  \(G\) is Newton's constant in units where \(\hbar=c=1\), and \(C\) is a constant whose value will turn out to be immaterial. Eq.~\eqn{shift} follows directly from boosting the Schwarzschild metric of the light particle to the light cone frame.

What it implies is that if a particle \(A\) with momentum \(p^\m=(0,0,p^-,0)\)  passes through a point \((0,0, 0,0)\), while an other particle \(B\) sits at a transverse location \((\tl x,0,0)\) as seen from that point, then the particle \(B\) will undergo a sudden displacement to the new position \((\tl x,\,-4Gp^-\log(|\tl x|/C),\,0) \). We see that, at small values of \(\tl x^2\), the spectator particle \(B\) is dragged along in the direction \(p^-\), as compared to another spectator particle further away (where \(|\tl x|\) is large).

If we assume that a black hole produced by one given state \(|\inn_0\ket\), upon its final explosion leads to a given final state \(|\out_0\ket\), one can now calculate the final state when a slight modification is brought about to the state \(|\inn\ket\). Let the modification consist of adding one light particle with momentum \(\d p^-\) entering the Rindler horizon at the transverse position \(\tl x\). All particles at the transverse position \(\tl x\,'\) in the final state \(|\out\ket\) are then dragged along such that their out-coordinate \(z^-\) is modified by an amount
	\be \d z^-=-4G \d p^-\log(|\tl x-\tl x\,'|/C)\ . \eel{oneparticle}
We write this modification as a property of the black hole scattering matrix:
	\be S\,|\inn_0\ket=|\out_0\ket\quad \ra\quad S|\inn_0+\d p^-(\tl x)\ket=e^{-ip^+_\out(\tl x\,')\d z^-}|\out_0\ket\ ,\eel{modif}
where we used the displacement operator \(e^{-ip^+(\tl x\,')}\) to describe a displacement at the transverse position \(\tl x\,'\).
	
	The modification \eqn{modif} can be repeated as many times as we wish, and this means that now we can reach any other initial state \(|\inn\ket\), when described by the distribution of the total momentum going in, \(p^-_\tot(\tl x)\) (as compared to the original initial state \(|\inn_0\ket\)), to find the new final state \(|\out\ket\) as a displacement of the original finite state \(|\out_0\ket\):
	\be\bra\out|S|\inn\ket=\bra \out_0|S|\inn_0\ket e^{4iG\int\dd^2\tl x\,'\log(|\tl x\,'-\tl x|/C)\,p^+_\out(\tl x\,')\,p^-_\inn(\tl x)}\ . \eel{smatrix1}
Note that, here, the operators \(p^-_\inn(\tl x)\) and \(p^+_\out(\tl x\,')\) both describe the total momenta of all in- and out going particles as distributions on the Rindler horizon. The important step to be taken now is to postulate that the entire Hilbert space of the in-particles is spanned by the function \(p^-_\inn(\tl x)\), and the black hole scattering matrix maps that Hilbert space onto the space of all particles going out, spanned by the function \(p^+_\out(\tl x\,')\). We arrive at the unitary scattering matrix \(S\):
	\be \bra p^+_\out|S|p^-_\inn\ket = \NN\,e^{4iG\int\dd^2\tl x\,'\log(|\tl x\,'-\tl x|/C)\,p^+_\out(\tl x\,')\,p^-_\inn(\tl x)}\ .	\eel{smatrix2}
There is only one unknown common factor, which we absorb in the normalization factor \(\NN\).

Since we only take their gravitational interactions into account in describing the back reaction, particles can \emph{only} be distinguished by their mass distributions. If other interactions are added, such as electro-magnetism, we will also be able to differentiate  particles further, for instance by their electric charges.	\def\intje{\textstyle{\int}} Therefore, Eq.~\eqn{smatrix2} must be regarded as an approximation. Concerns were expressed in Ref.\,\cite{Itzhaki} that some particles are also bent sideways so that these fall back into the horizon. In principle, this does not affect our analysis, since these particles do not occur in the final state at all, so that they are excluded in our picture. In any case this happens only at small transverse distance scales. We do expect that the transverse components of the gravitational force will invalidate our calculations at very high values of the transverse momenta. 
	
\newsecl{Algebra}{algebra}	
	From Eq.~\eqn{modif} and the expression \eqn{smatrix2} for the scattering matrix, we can now deduce the relations and commutation rules between the operators \(p^-_\inn(\tl x),\ p^+_\out(\tl x),\ z^+_\inn(\tl x)\) and \(z^-_\out(\tl x)\), where the latter can be regarded as the coordinates of in- and out going particles relative to the Rindler horizon:
	\be 		 z^-_\out(\tl x)&=&-4G\intje\dd^2\tl x\,' \,p^-_\inn(\tl x\,')\,\log(|\tl x-\tl x\,'|/C)\ ,\labell{zprel} \\[3pt]   
			z^+_\inn(\tl x)&=&4G\intje\dd^2\tl x\,' \,p^+_\out(\tl x\,')\,\log(|\tl x-\tl x\,'|/C)\ , \\ [3pt]      
			\,[ z^-_\out(\tl x),\,p^+_\out(\tl x\,') ]&=&[ z^+_\inn(\tl x),\,p^-_\inn(\tl x\,') ]\iss i\d^2(\tl x-\tl x\,')\ , \\[3pt]   
			\,[p^-_\inn(\tl x),\,p^+_\out(\tl x\,')]&=&-\fract i{8\pi G}\,\tl\pa^2\d^2(\tl x-\tl x\,')\ ,\labell{ppcom}\\[3pt] 
			\,[z^+_\inn(\tl x),\,z^-_\out(\tl x\,')]&=&	-4iG\log|\tl x-\tl x\,'|\ .\quad  \eel{zzcom}
In Eq.~\eqn{ppcom}, we used the fact that the logarithm obeys a Laplace equation,
	 \be \tl\pa^2\log|\tl x|=2\pi\d^2(\tl x)\ .\ee

It would have been more accurate to write the momentum distributions \(p^\pm(\tl x)\) as energy-momentum tensors \(T^{\pm,r}(\tl x,0,0)\), but we wished to emphasise that these are momenta, when integrated over the horizon.

So-far, we just reproduced the results of Refs.~\cite{GtHBH}. 	The algebra \eqn{zprel}---\eqn{zzcom} is quite different from the usual Fock space algebra for the elementary particles. In fact, it resembles a bit more the algebra of excited states of a closed string theory, but even that is not the same. 
It is therefore instructive to ask how one can decompose the new physical degrees of freedom into eigen modes.

\newsecl{Eigen modes}{eigen}	
The operators \(z^\pm(\tl x)\) and \(p^\pm(\tl x)\) form a linear set. Let us therefore consider plane waves in transverse space:
	\be p^\pm(\tl x)=\fract 1{2\pi}\int\dd^2\tl k\,\^p^\pm(\tl k)\, e^{i\tl k\cdot\tl x}\ ,\quad z^\pm(\tl x)=\fract 1{2\pi}\int\dd^2\tl k\,\^z^\pm(\tl k)\, e^{i\tl k\cdot\tl x}\ , \eel{transvwaves}
to find that the \(\tl k\)-waves indeed decouple:
	\be & \tl k^2\,\^z^-_\out(\tl k)=8\pi G\,\^p^-_\inn(\tl k)\ , \quad \tl k^2\,\^z^+_\inn(\tl k)=-8\pi G\,\^p^+_\out(\tl k)\ , &\labell{zpk}\\[3pt]
	&[\^z^-_\out(\tl k),\,\^p^+_\out(\tl k')]=[\^z^+_\inn(\tl k),\,\^p^-_\inn(\tl k')]=i\d^2(\tl k-\tl k')\ ,& \eel{zpk2}
etc. Eqs.~\eqn{zpk} can be seen as boundary conditions on the horizon; the waves moving in are transformed into waves going out. 

	Note that the waves \eqn{transvwaves} on Rindler space should not be interpreted as particles; they each may consist of many particles, and we should only consider their real parts as physical objects. A single physical particle is typically represented by a Dirac delta distribution, see Eq.~\eqn{oneparticle}, where we have \(p^-_\inn(\tl x\,')=\d p^-\,\d^2(\tl x\,'-\tl x)\). Thus, at this stage of our presentation, the momenta \(p^\pm\) are to be interpreted quite differently from the transverse wave number variables \(\tl k\) on Rindler space. The waves \(\^z^\pm(\tl k)\) should be split into sines and cosines, which each describe a wavy displacement operator on the horizon.

The above boundary equations are not very intuitive. This is because the time evolution of the free waves are in the form of scaling equations. In terms of the Rindler time coordinate \(\t\) we have: \(p^-_\inn(\t)=e^\t\,p^-_\inn(0)\,,\ \ p^+_\out(\t)=e^{-\t}\,p^+_\out(0)\), etc. Also, \(p^\pm\) and \(z^\pm\) are operators, not amplitudes.

What happens becomes more transparent if we replace the Rindler space coordinates  \(z\)  and the momenta \(p\) by tortoise (Eddington-Finkelstein) coordinates \(\r\) and \(\w\), Then, however, we must take into account that both the \(z\) and the \(p\) coordinates may be positive or negative. Therefore, we also introduce signs \(\a=\pm 1\) and \(\b=\pm 1\):
	\be z^+_\inn\equiv \a\,\ex\r\ ,\quad p^-_\inn\equiv \b\,\ex\w\ . \eel{tortoise}

How are the operators \(\b\) and \(\w\) related to \(\a\) and \(\r\)? This should be easy. The transformations \eqn{tortoise} imply for the wave functions,
	\be \j(\a\,\ex{\r})=\ex{-\half\r}\,\vv(\a,\,\r) \ , \eel{expo1}
where \(\j\) is the amplitude in the variable \(z^+_\inn(\tl k)\) for some fixed \(\tl k\), while \(\vv\) is a wave function in the new tortoise coordinate \(\r\); and \(\j\) is normalized in the \(z\) variable while \(\vv\) is normalized in the \(\r\) variable, which required the factor \(e^{\,\!-\r/2}\). Similarly, we write in momentum space,
	\be  \^\j(\b\,\ex{\w})=\ex{-\half\w}\,\^\vv(\b,\,\w) 	\eel{expo2}
(Note, that the caret \((\,\!\^\ )\) is now used for Fourier transformation in \(z^+\) space, as the caret in Eqs.~\eqn{zpk} and \eqn{zpk2} is no longer needed; from now on we look at just one value for the transverse wave function \(\tl k\)).

Now,  the relation between the wave function \(\j\) and its Fourier transform \(\^\j\) is
	\be &\ds\vv(\a,\,\r)=\fract 1{\sqrt{2\pi}}e^{\halfje\r}\int_{-\infty}^\infty\dd p^-_\inn e^{i p^-_\inn z^+_\inn}\,\^\j(p^-_\inn)\ =& \nn
		&\ds\fract 1{\sqrt{2\pi}}\sum_{\b=\pm}\int_{-\infty}^\infty\ex{\half(\r+\w)}\dd\w\,\ex{\a\b\,ie^{\r+\w}}\,\^\vv(\b,\w)\ = &\\
		&\ds\fract 1{\sqrt{2\pi}}\sum_{\b=\pm}\int_{-\infty}^\infty\dd u\,A(\a\b,\,u)\^\vv(\b,\,u-\r)\ ;\quad A(\s,\,u)=\ex{\half u+
		i\s e^u}\ . &\nm \ee
We see the emergence of a matrix \(A(\a,\b)=({A(+)\ A(-)\atop A(-)\ A(+)})\), which is easy to diagonalise; simply write for  \(\vv\),
	\be \vv(+,\w)+\vv(-,\w)=\vv_1(\w)\ , \quad \vv(+,\w)-\vv(-,\w)=\vv_2(\w)\ , \ee
and similarly for  \(\^\vv(\b,\,\w)\) and \(A(\s,\,u)\), to find the diagonalised expressions
	\be &\ds\vv_i(\r)=\int_{-\infty}^\infty\dd u\,A_i(u)\^\vv_i(u-\r)\qquad(i=1,2)\ ,& \labell{expfourier} \\
	&\ds A_1(u)=\sqrt{\fract 2{\pi}}\,e^{\halfje u}\cos(e^u)\ ,\quad A_2(u)=i\sqrt{\fract 2{\pi}}\,e^{\halfje u}\sin(e^u)\ .&\ee
Similarly,
	\be \^\vv_i(\w)=\int_{-\infty}^\infty\dd u\,A^*_i(u)\vv_i(u-\w)\ .\eel{expfourier2}
	
We can write the boundary conditions \eqn{zpk} as
	\be z^-_\out=\l\,p^-_\inn\ ,\quad p^+_\out=-\fract 1\l z^+_\inn \ , \ee
where we take for granted that we take the Fourier coefficient with Rindler wave vector \(\tl k\), and define \(\l=8\pi G/\tl k^2\). And now consider \(p^-\) as the Fourier variable to \(z^+\). Applying Eqs.~\eqn{expfourier} and \eqn{expfourier2} then gives
	\be \vv^\out_i(\r)=\int_{-\infty}^\infty\dd u\, A^*_i(u)\,\vv^\inn_i(u+\log\l-\r)\ . \eel{inouttortoise}

In the tortoise coordinates \(\r\) and \(\w\), we see that the waves move in and out with velocity one. This means that an in going wave \(\Psi_i^\inn\) and an out going wave \(\Psi_i^\out\) can be written in terms of plane waves as 
	\be \Psi_i^\inn(\r,\,\t)=\int\dd\k \tl\Psi_i^\inn(\k)\,\ex{i\k(-\r-\t)}\ , \\
	 \Psi_i^\out(\r,\,\t)=\int \dd\k \tl\Psi_i^\out(\k)\,\ex{i\k(\r-\t)}\ ,  \ee
where \(\t\) is the Rindler time variable, \(\k\) the Fourier parameter in the tortoise coordinates, and \(\tl\Psi\) now denotes the Fourier coefficient in the tortoise coordinates (we use the tilde  \((\,\tl{ }\,)\) rather than the caret because, again, this is a different Fourier transformation than the one in Eqs.~\eqn{expo2}---\eqn{expfourier2}).

This then yields the final part of our diagonalization process, since these plane waves also diagonalise Eq.~\eqn{inouttortoise}:
	\be \tl\Psi^\out_i(\k)=\tl A^*_i(\k)\,e^{-i\k\log\l}\,\tl\Psi_i^\inn(\k)\ . \eel{bounce}
The coefficients \(\tl A_i(\k)\) are the Fourier coefficients of \(A_i(u)\) and can be given in closed form. By contour integration, one derives:
	\be  &\ds \tl A_1(\k)=\fract 1{\sqrt\pi}\G(\half+i\k)(\cosh{\pi\k\over 2}+i\sinh{\pi\k\over 2})\ ,& \\[3pt]
		&\ds \tl A_2(\k)=\fract 1{\sqrt\pi}\G(\half+i\k)(\sinh{\pi\k\over 2}+i\cosh{\pi\k\over 2})\  .& \ee
These coefficients have norm one. This is verified using 
	\be \cosh^2(x)+\sinh^2(x)=\cosh(2x)\quad\hbox{and}\quad\G(\half+i\k)\G(\half-i\k)={\pi\over\cosh\pi\k}\ . \ee

This completes our diagonalization process. We see that Eqs.~\eqn{inouttortoise} and \eqn{bounce} can be seen as a real bounce against the horizon. The information is passed on from the in-going to the out-going particles. We do emphasise that in- and out-going particles were not assumed to affect the metric of the horizon, which is fine as long as they do not pass by one  another at distances comparable to the Planck length or shorter; in that case, the gravitational effect of the transverse moments must be taken into account. For the rest, no other assumptions have been made than that the gravitational fields of in- and out going particles should not be ignored. This must be accurate as long as we keep the transverse distances on the horizon large compared to the Planck length.

It is also important to emphasise that, even though we describe modes of infalling matter that ``bounce back against the horizon", these bounces only refer to the information our particles are carrying, while the particles will continue their way falling inwards as seen by a co-moving observer. In accordance with the notion of Black Hole Complementarity, an infalling observer only sees matter going in all the way, and nothing of the Hawking matter being re-emitted, since that is seen as pure vacuum by this observer. Rather that stating that this would violate no-cloning theorems, we believe that this situation is asking for a more delicate quantum formalism.

\newsecl{The black hole entropy}{entropy}
	One could try to compute the black hole entropy from the contributions of these reflecting modes. For each mode, the result is finite. The entropy is found from the free energy \(F\), which is defined by
	\be e^{-\b F}=\sum_{\mathrm{states}}e^{-\b E}=\sum_{\k_n}e^{-\b \k_n}\ , \eel{freeF}
where \(\b\) is now taken to be the inverse of the Hawking temperature, later to be substituted by its (presumable)  value \(2\pi\).

The energy modes \(\k_n\) are derived by assuming that Eq.~\eqn{bounce} provides for the proper boundary condition near the horizon; the wave bounces at the value \(\r_0\)  of \(\r\) where \(\Psi_i^\out(\r,\,\t)=\Psi_i^\inn(\r,\,\t)\), which is where
	\be   e^{i\k(\r-\t)}\,\tl A_i^*(\k) \, e^{-i\k\log\l}&=& e^{i\k(-\r-\t)}\ , \quad \hbox{so that} \nn
		i\k(2\r_0-\log\l)-i\a_i(\k)&=&0\ ,  \ee
where the angles \(\a_i(\k)\) are the arguments of the coefficients \(\tl A_i(\k)\):
	\be \tl A_i(\k)\equiv e^{i\a_i(\k)}\ . \ee
Assuming a box with outer edge \(\r=\r_1=\log\L\), one finds that the values for \(\k_n\) must obey 
	\be\pi n\iss\k_n(\r_1-\r_0)\iss\k_n(\log\L-\half\log\l)-\half\a_i(\k_n)\ .				\ee

A rough estimate for \(\a_i(\k)\) is obtained by applying Stirling for large \(\k\):
	\be \tl A_i(\k)=e^{i\a_i(\k)} \ \ra \ \ex{i\k(\log\k-1)+\pi i/4}\ . \ee 
Taking \(L\) sufficiently large, we elaborate Eq.~\eqn{freeF}:
	\be e^{-\b F}&=&{\textstyle\sum_{i=1}^2}\int_0^\infty\dd n\,e^{-\b\k_n}\iss{\textstyle\sum_{i=1}^2} \int_0^\infty e^{-\b \k}\,\dd\k{\dd n\over\dd \k}\ \approx \nn
		&\approx&	\fract 2 \pi\int_0^\infty e^{-\b\k}\,\dd\k\,(\log\L-\half\log\l-\half\log\k)\ = \nm\\[3pt]
		&=& \fract 1 {
		\pi\b}(2\log\L+\log\b+\g-\log\l)\ ,	\eel{freeF2}
where \(\g\) is Euler's constant. The cut-off \(\L\) refers to the edges of the box in which we keep the black hole, so \(\log\L\) in Eq.~\eqn{freeF2} merely refers to the contribution of Hawking radiation in the empty space far from the black hole.

Using the thermodynamical equations 
	\be U=\fract\pa{\pa\b}(\b F)\ ,\quad S=\b(U-F)\ , \eel{thermo}
one can derive the contribution of each mode with transverse wave number \(\tl k\) to the total entropy. 

\newsecl{Duscussion}{disc}
Note, that we did not apply second quantization, such as in Ref.\,\cite{GtHBH0}, since now we are not dealing with a quantum field theory. At every value of \(\tl k\), there are exactly two wave functions \(\Psi(\pm,\r,\t)\) (one at each side of the horizon, which mix).

The expression we obtained must now be summed over the values \(\tl k\). If we take these to describe a finite part of the black hole horizon area, we see that the summed expression will be proportional to the area, as expected, but the sum diverges quadratically for large \(\tl k\). Since \(\l=8\pi G/\tl k^2\), Eq.~\eqn{freeF2} does depend on \(\tl k\), but too weakly, even slowly increasing for large \(\tl k\).

The explanation for this divergence is that, as noted at the end of  section~\ref{eigen}, our expressions are inaccurate at very large \(\tl k\), where transverse gravitational forces should be taken into account.

It is not easy to correct for this shortcoming, but we can guess how one ought to proceed. It was remarked already in Refs.\,\cite{GtHBH}, that the algebraic expressions we obtain on the 2-dimensional horizon, take the form of functional integrals very much resembling those of string theory. We did treat the transverse position variables \(\tl x\) and wave number variables \(\tl k\) very differently from the longitudinal variables \(z^\pm\) and \(p^\pm\), but it is clear that we are dealing with the full expressions of an \(S_2\) sphere. This sphere should be given two arbitrary coordinates \(\tl\s=(\s^1,\,\s^2)\), after which these should be fixed by a gauge condition relating them to the transverse coordinates \(\tl x\). We took \(\tl \s=\tl x\), but apparently this fails when the longitudinal variables fluctuate too wildly.

It was also observed that the original in-going and out-going particles with which we started, produce vertex insertions as in a string world sheet, as if all particles considered should be regarded as closed string loops. It all takes the form of a string theory. Strings were not put in, however, rather, they come out as inevitable objects! But beware, these are not ``ordinary" strings. The black hole horizon is the string world sheet.  If ordinary strings were to be Wick rotated to form space-like string world sheets, all factors \(i\) would disappear from the action, whereas our expressions are still in the complex plane, \emph{as if the string slope parameter \(\a'\) should have the purely imaginary value \(4 Gi\).} In most string treatments of black holes, the string world sheets are assumed to be in the longitudinal direction, that is, the world sheets are take to be orthogonal or dual to the horizon.

Our analysis appears to be closely related to ideas using the BMS approach\,\cite{Hawking1}, although there the emphasis seems to be specially on the in- and out-going gravitational waves, while we focus on all particle types entering or leaving the black hole. Secondly, although it is clearly of importance to consider measurements made at \(\mathfrak I^+\) and \(\mathfrak I^-\), we attribute the black hole properties to the immediate surroundings of the future and past event horizon. Also, one may note that both approaches now focus on light-like geodesics, which justifies attempts to employ conformally invariant descriptions of quantum gravity.

\end{document}